\documentclass[prb,aps,twocolumn]{revtex4}
\usepackage{epsf}
\usepackage{epsfig}
\usepackage{hyperref}
\usepackage{amsmath}
\begin{document}

\title{
Antiferromagnetic Order and Phase Coexistence \\
in Antisite Disordered Double Perovskites
}

\author{Viveka Nand Singh and Pinaki Majumdar}

\affiliation{Harish-Chandra  Research Institute,
 Chhatnag Road, Jhusi, Allahabad 211019, India}

\date{9 Sep, 2010}

\begin{abstract}
In addition to the well known ferromagnetism, double perovskites are also 
expected to exhibit antiferromagnetic (AF) order driven by electron 
delocalisation. This has been seen in model Hamiltonian studies  and confirmed 
via {\it ab initio} calculations. The AF phases should occur, for example, on 
sufficient electron doping of materials like Sr$_2$FeMoO$_6$ (SFMO) via La 
substitution for Sr. Clear experimental indication of such AF order is limited, 
possibly because of increase in antisite disorder with La doping on SFMO, 
although intriguing signatures of non ferromagnetic behaviour are seen. We 
study the survival of electronically driven antiferromagnetism in the presence 
of spatially correlated antisite disorder and extract the signals in magnetism 
and transport.  We discover that A and G type AF order, that is predicted in 
the clean limit, is actually suppressed {\it less strongly} than ferromagnetism 
by antisite disorder. The AF phases are metallic, and, remarkably, more conducting 
than the ferromagnet for similar antisite disorder.  We also highlight the phase 
coexistence window that connects the ferromagnetic regime to the A type 
antiferromagnetic phase.

\end{abstract}

\maketitle

\section{Introduction}
The double perovskites \cite{dp-nat,dd-rev,serrate-rev}
(DP) are materials of the form ABO$_3$AB$^{\prime}$O$_3$.
The B and B$^{\prime}$ are usually transition metal ions, while A is a 
rare earth or alkaline earth. The itinerant electrons in this
system are coupled strongly to the 
transition metal magnetic moments, 
which act as core spins, and the magnetic order is 
driven \cite{theor-abin-dd,theor-millis}
by minimisation of the electronic energy, 
rather than a short range interaction between the moments. 
Much of the excitement in the double perovskites has been
due to the high ferromagnetic $T_c$, noticeable magnetoresistance,
and the possibility of spin polarised conduction in materials
like Sr$_2$FeMoO$_6$ (SFMO).
  
While the ferromagnetism is certainly useful, one may wonder
about the occurence of other kinds of magnetic order in these
materials.  For example, even `simple perovskite' transition metal 
oxides, the cuprates, manganites,
or cobaltates, have a rich phase diagram \cite{dag-sc-rev}, 
with a strong dependence
on the doping level. The manganites, for instance, exhibit
not just ferromagnetism, but also 
`CE' magnetic order and A, C, and G type
antiferromagnetic (AF) phases \cite{tok-mang-rev}, 
depending on the hole doping level. 
It is interesting to explore if non-ferromagnetic 
ordered states are possible in the double perovskites as
well.

A study of the model Hamiltonian of these materials shows that 
at low electron density a ferromagnetic (FM) alignment of the core
spins is favoured since it leads to the maximum bandwith. However,
at suffciently large band filling, antiferromagnetic (AF) states
with A or G type order (in two dimensions) successively become 
favoured \cite{ps-pm-scr,dp-af-tsd,dp-af-tm}. 
While these spin configurations lead to smaller electronic 
bandwidth they have a higher density of band edge states compared
to the ferromagnet.

We have studied these magnetic states in two
dimensions \cite{ps-pm-scr} and three dimensions \cite{dp-af-rajr}
 elsewhere.
{\it Ab initio} calculations using simple collinear arrangement
of the `core spins' predict similar results \cite{dp-af-tsd,dp-af-tm}
in the context of real materials.
Nevertheless, electronically driven AF order continues to be
elusive experimentally.
One can anticipate two reasons: (i)~the AF phase occurs at
(high) electron densities which have not been probed yet, or
(ii)~increasing electron density leads to a rapid growth
in antisite disorder (ASD), strongly suppressing any signature
of long range AF order.

The theoretical effort till now has focused on
AF order in the ideal structurally
ordered background, where the B and B$^{\prime}$ 
ions of A$_2$BB$^{\prime}$O$_6$ alternate
along each axis.
The situation in the real material is far from
ideal, and the B, B$^{\prime}$ alternation is interrupted by an
antiphase boundary (APB)  \cite{asd-tok-dom,asd-dd-dom}
involving  BB or B$^{\prime}$B$^{\prime}$ nearest neighbours. 
This leads to a pattern of structural
domains. The ASD
not only destroys periodicity but also brings into play
an additional AF superexchange coupling when two B ions (like Fe) 
adjoin each other.  If this superexchange
scale is sufficiently large, the BB antiphase boundary also
acts as a magnetic domain wall \cite{asd-tok-dom} (MDW).
The effect of such AF superexchange in the double perovskite
ferromagnets is well known. There it leads to magnetic 
domain formation,
suppressing the magnetisation,
and large enhancement of the resistivity \cite{expt-asd-huang,expt-asd-nav}.
 The interplay of antisite disorder and AF superexchange with
{\it itinerant electron antiferromagnetism} is unexplored.

This paper provides a systematic exploration of the effect
of increasing antisite disorder 
on the AF phases in a two dimensional (2D) double 
perovskite model. We work in 2D for ease of visualisation 
and to access large system size.

The phase diagram mapping out the occurence of AF phases in
the `clean' limit has been established earlier \cite{ps-pm-scr}.
Here we focus on 
a couple of electron densities, one each in the `A type'
and `G type' window, respectively, and study the magnetic
order and transport for increasing antisite disorder.
Our principal results are the following. (i)~The suppression 
of magnetic order with ASD is {\it slower} in the AF phases compared
to the ferromagnet, and even with $\sim 30\%$ mislocation
there are clear signatures of long range order. 
(ii)~The $T_c$ is not significantly affected,
until strong disorder, although the `transition' is broadened.
(iii)~Both AF phases are metallic, and the increase in their
resistivity with ASD is {\it weaker} than in the ferromagnet.
(iv)~The phase separation regime between the FM and A type phase
is converted to a {\it phase coexistence} window in the presence
of moderate ASD, and one observes the occurence, locally, of
both kinds of order in the same sample.

The paper is organised as follows. We first quickly review the
results on AF phases obtained in model studies and {\it ab initio}
calculations, and follow it up with a survey of the few experiments
in this regime. We then discuss the model and method. This is
followed by a discussion of our results on
the AF order in the background of antisite disorder, with
focus on the spatial correlations and the transport properties
of the system. We conclude with a discussion of the phase coexistence
region that intervenes between the ferromagnetic and A type phase,
and is likely to be of relevance in La doped SFMO.

\section{Studies on AF order}

Early studies using model Hamiltonians for double perovskites had
observed the instability \cite{theor-millis} 
of the ferromagnetic state, without
exploring the competing phase  that emerges.
A subsequent variational study \cite{theor-asd1} 
did identify non-ferromagnetic phases. 
More recent studies using both simple models \cite{ps-pm-scr}
 and realistic DFT calculations \cite{dp-af-tsd,dp-af-tm}
indicate that the ferromagnet becomes unstable to an A type phase
on increasing electron density. In 
Sr$_{2-x}$La$_x$FeMoO$_6$, for example, this is expected to happen
for $x \gtrsim 1$. 
The DFT studies have employed supercells for a few commensurate doping
levels. Using a three band model Hamiltonian, with parameters 
inferred from the DFT, the authors 
have explored \cite{dp-af-tsd} 
a more continuous 
variation of La doping level and confirmed the DFT trends.
The crossover to a non FM ordered state is, therefore,
not an artifact of
a single band model or two dimensionality that earlier studies
employed.
As for the effect of ASD on non ferromagnetic phases, we are
aware of only one study involving uncorrelated antisite defects
\cite{theor-asd1}. It is more focused on the doping dependence, 
and explores mainly the  magnetism, but the trends are
consistent with what we observe here.

Samples have indeed been synthesised with large La doping on SFMO
\cite{af-expt1,af-expt2}. 
The main observations are (i)~a suppression \cite{af-expt1} of
the low field magnetisation with increasing $x$, and (ii)~a large
difference \cite{af-expt2}
 between the field cooled (FC) and zero field cooled
(ZFC) response. There is unfortunately no detailed understanding 
of the ASD in these samples yet (for example via XAFS), 
or data on resistivity and magnetoresistance.

\section{Model and method}

The model we study has been presented earlier in the context of
the ferromagnetic phase, so we only provide a brief description 
for completeness.

The information about the B, B$^{\prime}$ structural motif is encoded in
a binary variable $\eta_i$, with
$\eta_i =1$ for B sites and $\eta_i =0$ for B$^{\prime}$ sites. 
In the structurally ordered DP the $\eta_i$ 
alternate along each axis. We will consider progressively
`disordered' configurations, generated through an annealing process.
For any specified $\{\eta\}$ background the model has the form:
\begin{eqnarray}
H&=&
\epsilon_{B}\sum_{i \sigma}\eta_{i}f_{i\sigma}^{\dagger}f_{i\sigma}+
\epsilon_{B^{\prime}}\sum_{i \sigma}\left(1-\eta_{i}\right)  
m_{i\sigma}^{\dagger}m_{i\sigma}\nonumber\\
&& + H_{kin} \{ \eta\} 
+J\sum_{ i \alpha \beta}\eta_{i} {\bf S}_{i}\cdot 
f_{i\alpha}^{\dagger}\overrightarrow{\sigma}_{\alpha\beta}f_{i\beta}
- \mu {\hat N} \nonumber\\
&&+J_{AF}\sum_{\left\langle i,j\right\rangle }\eta_{i}
\eta_{j} {\bf S}_i\cdot{\bf S}_j
\end{eqnarray}

$f$ is the electron operator on the magnetic 
B site and $m$ is the operator on the non-magnetic 
B$^{\prime}$  site. $\epsilon_B$ and 
$\epsilon_{B^{\prime}}$ are onsite energies, 
at the B and B$^{\prime}$ sites respectively.
$\epsilon_{B} - \epsilon_{B^{\prime}} $ 
is a  `charge transfer' energy.
$H_{kin}$ is the electron hopping term:
$-t_1\sum_{\langle i,j \rangle  \sigma}\eta_i \eta_j
f_{i\sigma}^{\dagger}f_{j\sigma} $ 
$ -t_2\sum_{\langle i,j \rangle \sigma}
( 1-\eta_i) (1-\eta_j) 
m_{i\sigma}^{\dagger}m_{j\sigma} $ 
$ -t_3\sum_{{ \langle i,j \rangle} \sigma}
( \eta_i+\eta_j-2\eta_i \eta_j)
( f_{i\sigma}^{\dagger}m_{j\sigma}+h.c)$.
The $t$'s are all nearest neighbour hopping amplitudes,
and for simplicity we set $t_1=t_2=t_3=t$ here.
 $ {\bf S}_i$ is
the core spin on the site ${\bf R}_i$, with
$\vert {\bf S}_i \vert =1$. 
$J$ is the Hund's
coupling on the B sites, and we use $J/t \gg 1$. When the up-spin core levels
are fully filled, {\it e.g.}, Fe $t_{2g\uparrow}$ and $e_{g\uparrow}$ in
SFMO, the conduction electron is forced
to be {\it antiparallel} to the core spin. We have used $J>0$ to
model this situation. For the present study we have set 
\cite{footnote-b-bprime} the
{\it effective } level difference  $
\epsilon_B -J/2 - \epsilon_{B^{\prime}} =0$.
The chemical potential $\mu$ is used to control the electron density,
and ${\hat N}$ is the total electron density operator.
We study one point in the A type window, and another in the G type
window. We also explore the transition from FM to A type order.

When two magnetic atoms can be on neighbouring sites,
we also have to consider the {\it antiferromagnetic}  coupling 
$J_{AF}$ between nearest neighbour B sites. 

It has been well established now that the antisite disorder
does not involve `random' exchange of B and B$^{\prime}$ ions, but follows
a correlated pattern \cite{asd-tok-dom,asd-dd-dom}. 
A periodic B-B$^{\prime}$ pattern is interrupted by
a line-like defect, where BB or 
B$^{\prime}$B$^{\prime}$ adjoin each other, and beyond
this boundary one obtains another domain but with a phase slippage
with respect to the first one. The line defect is an antiphase
boundary.
If a fraction $x$ of atoms are `mislocated' with 
respect to the ideal ordered structure, these
atoms themselves are organised into domains, and do not
act as simple `point defects'. 

We generate the correlated patterns by studying a lattice gas
model \cite{ps-pm-asd}, 
with poor annealing to prevent long range order. Briefly,
the $\{ \eta_i \}$ configuration arise from:
\begin{equation}
H_{eff} \{ \eta \} = -V \sum_{\langle ij\rangle} \eta_i (1 - \eta_j)
\end{equation}
The $\eta$ are coupled only between nearest neighbour sites.
The ground state in this model corresponds to $\eta =1,0,1,0,..$
along each axis, {\it i.e}, B, B$^{\prime}$, B, B$^{\prime}$..
$V > 0$ is a measure of the ordering tendency, and sets the
temperature for long range B-B$^{\prime}$ order if the system were allowed
to equilibriate. 

\begin{figure}[t]
\centerline{
\includegraphics[width=2.2cm,height=2.2cm,angle=-90,clip=true]{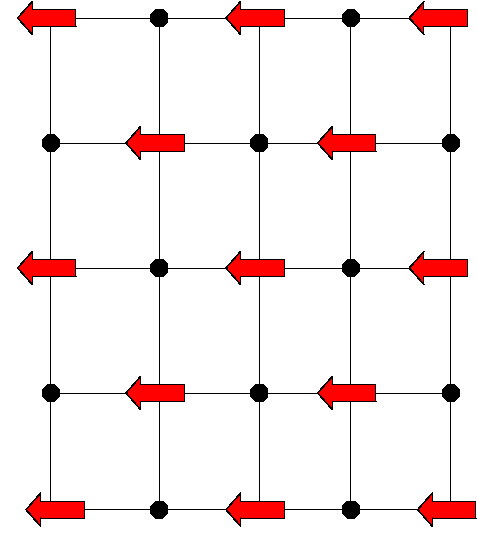}
\hspace{.5cm}
\includegraphics[width=2.2cm,height=2.2cm,angle=-90,clip=true]{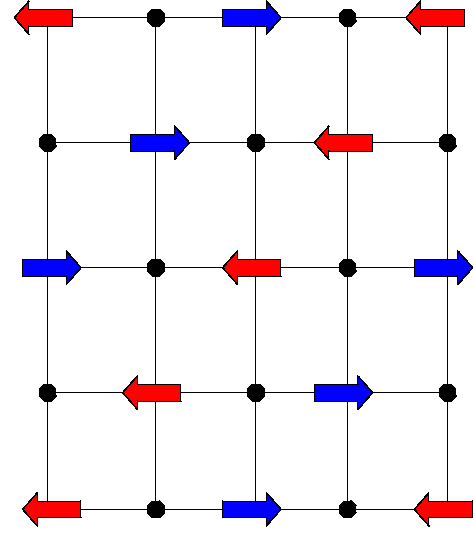}
\hspace{.5cm}
\includegraphics[width=2.2cm,height=2.2cm,angle=-90,clip=true]{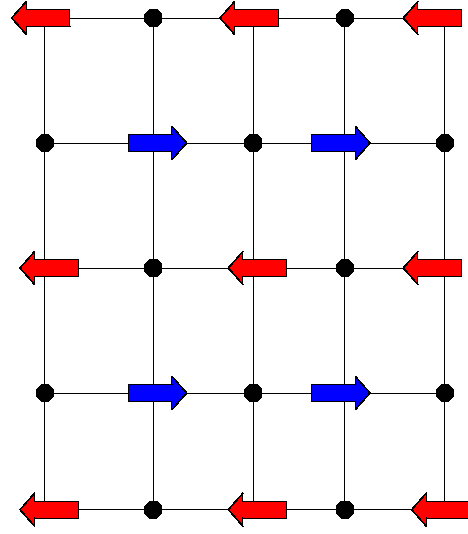}
}
\vspace{.8cm}
\centerline{
\includegraphics[width=6.5cm,height=4.5cm,angle=0,clip=true]{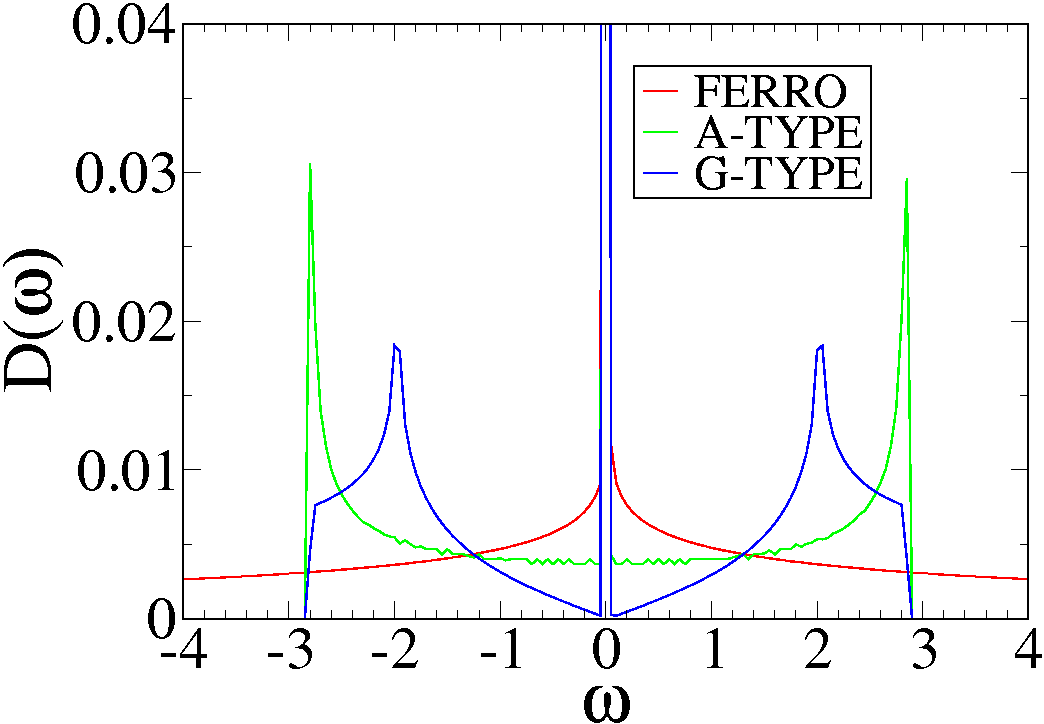}
}
\caption{Top: The three magnetic phases in the structurally ordered
2D double perovskite model.
Left: ferromagnetic (FM), center: 
A type antiferromagnet, right:  G type antiferromagnet.
These occur with increasing
electron density. The moments are on the B sites.
We have not shown the induced moments on the B$^{\prime}$ sites.
Bottom: Electronic density of states
for the ferromagnetic, A and G type ordered phases in the
structurally ordered background.}
\end{figure}

We used different `annealing time', 
$\tau_{ann}$, and annealing temperature,
$T_{ann}$,
to generate the ASD configurations.
The ASD structures are characterised in terms of the
following indicators:
$(i)$ The structural `order parameter'  $S=1-2x$,
where $x$ is the fraction of
B (or B$^{\prime}$) atoms that are  on the wrong sublattice.
$(ii)$~The degree of short range order, characterised by the probability,
$p$, of having nearest neighbour pairs that are B-B$^{\prime}$. 
$(iii)$~The correlation length $\xi$ in these structures,
computed from the width of the ordering peak.

For a given $\{\eta_i\}$
configuration, generated at some $T_{ann}$ and $\tau_{ann}$,
we solve for the magnetic
properties, and electronic properties averaged over equilibrium
magnetic configurations, via a cluster based Monte Carlo technique
\cite{tca}.
Electronic properties are calculated after 
equilibration by diagonalizing the full system.

The electronic conductivity is calculated using the
Kubo formula, computing the
matrix elements of the current operator. The `dc conductivity'
is the low frequency average, $(1/{\Delta \omega}) 
\int_{0}^{\Delta\omega}\sigma(\omega)d\omega $
of the optical conductivity.
This is averaged over thermal configurations and disorder, as 
appropriate. We use $\Delta \omega = 0.05t$.

\section{Results}

\begin{figure}[b]
\centerline{
\includegraphics[width=6.0cm,height=4.0cm,angle=0,clip=true]{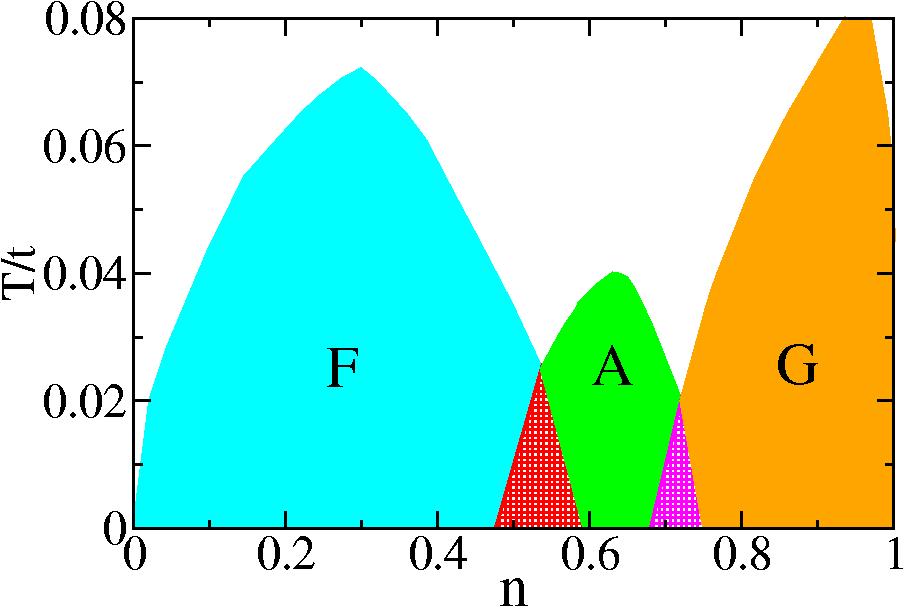}
}
\caption{Colour online: 
Phase diagram for the non disordered double perovskite. We
only show the region $n=[0,1]$. From $n=1$ to $n=2$ one populates
the non-dispersive B$^{\prime}$ level, and the magnetic state is G type. The
$n=[2,3]$ window is a symmetric version of the $n=[1,2]$ region. 
The regions between the phases indicate phase separation.
The results are obtained via MC on a $40 \times 40$ lattice.}
\end{figure}

\subsection{The clean limit}

In the absence of antisite disorder there are no AF superexchange
interactions in the system, and the magnetic order is decided by
minimisation of the electronic energy. 
We reproduce the model for the ordered DP, below, for ease of
reference:
\begin{eqnarray}
H_{ord} & =&~~~ \epsilon_{B}\sum_{i\in B, \sigma}f_{i\sigma}^{\dagger}f_{i\sigma}+
\epsilon_{B^{\prime}}\sum_{i\in B^{\prime} \sigma}m_{i\sigma}^{\dagger}m_{i\sigma}
-\mu {\hat N}  \cr
&&~  -t\sum_{<ij>\sigma}f_{i\sigma}^{\dagger}m_{j\sigma}
+ J\sum_{i\in A, \alpha \beta} {\bf S}_{i} \cdot
f_{i\alpha}^{\dagger}\vec{\sigma}_{\alpha\beta}f_{i\beta}
\nonumber
\end{eqnarray}
For $J/t \gg 1$ this model supports 
three collinear phases
(in 2D). The FM state gives way to A type (line like) order with
increasing electron density, and finally to a G type state. These have
been discussed earlier \cite{ps-pm-scr}, 
we reproduce the magnetic configurations, top row in Fig.1, 
and the electronic density of states below. We have set
$ (\epsilon_B - J/2) - \epsilon_{B^{\prime}} =0$.

\begin{figure}[t]
\includegraphics[width=7.5cm,height=10.0cm,angle=0]{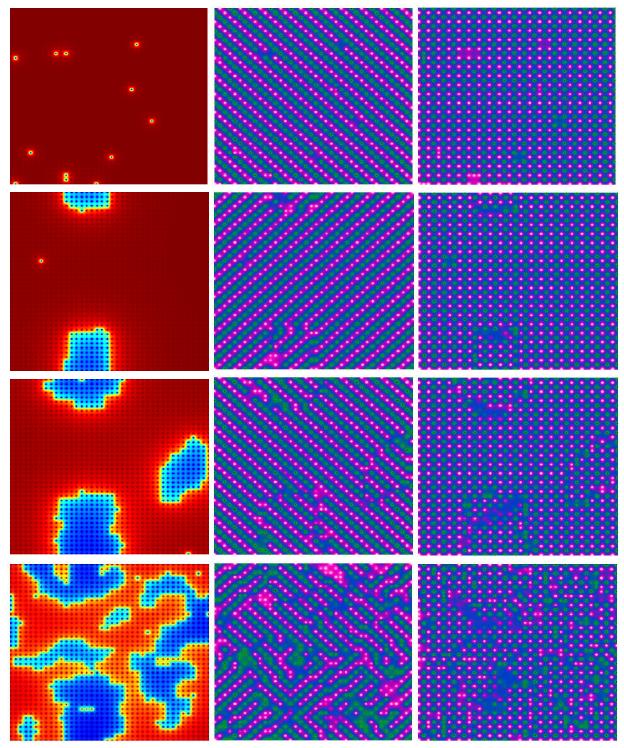}
\caption{Colour online: Antiphase domains and
corresponding antiferromagnetic phases. The left column shows the
domain pattern in the ASD background, with structural order
parameter $S=0.98,~0.76,~0.50,~0.08$ as we move from top to bottom.
The middle column shows the A type AF phase on this structural motif,
while the right column shows the G type phase.
The magnetic correlations are characterised via the
overlap factor
$g_i = {\bf S}_0.{\bf S}_i$, where ${\bf S}_0$ is the left lower corner spin
in the lattice.
}
\end{figure}


The FM state is preferred at low $n$, since it has
the largest bandwidth. The A type state has lower bandwidth, but
with large density of states 
near the band edge. The FM becomes unstable at
$n \sim 0.45$. 
The A type state
is stable for $n \ge 0.58$, and between these we have a 
phase separation (PS)  window. Similarly the A to G transition involves
a PS window. The PS windows narrow with increasing $T$ and 
vanish as $T \rightarrow T_c$.
The thermal transitions and the PS windows are shown
in Fig.2.

Broadly, the task would be to extend this phase diagram to finite
antisite disorder. Instead of attempting to map out the disorder
dependence at {\it all densities} we choose two representative densities,
$n \sim 0.65 $ in the A type window, and $n \sim 0.95$ in the 
G type region, to clarify
the impact of disorder. We also explore the effect of ASD on the 
phase separation window since it would be encountered in any attempt
to electron dope the ferromagnet.

\subsection{Disorder configurations}

We study four families, with progressively increasing 
antisite disorder. Each family arises from annealing the
lattice gas model for some time $\tau_{ann}$ at temperature
$T_{ann}$. 
A representative configuration from each family is shown in the 
first column in Fig.3.
They have a fraction of mislocated sites:
$x = 0.01,~0.12,~0.25,~0.46$. 
The fraction of course varies somewhat 
from sample to sample within each
family. Our `disorder average' for magnetic and electronic properties
is performed typically over 10 configurations within each family.

The structural `order parameter' 
 $S= 1 - 2x$ is $0.98,~0.76,~0.50,~0.08$
for these four families. Since the configurations emerge from an
annealing process, the spatial correlations in the structure are
much stronger than the $S$ value would suggest. In the absence of
information about spatial correlations, the probability $p$ of having
a BB$^{\prime}$ nearest neighbour bond is 
$p = x^2 + (1-x)^2 =
(1/2) (1 + S^2)$. The two terms, $x^2$ and $(1-x)^2$, arise
from having both atoms on the `wrong' sublattice or
`right' sublattice respectively. For the $S$ values in
our configurations these numbers are 
$\sim 0.98,~0.79,~0.63,~0.50$. 
In our case the structures are spatially correlated, so
the probability of BB$^{\prime}$ nearest neighbours is much higher 
than the estimates, above,  from an `uncorrelated distribution'. 
By analysing
our configurations we obtain:
$p \sim 0.98,~0.97,~0.95,~0.86$, indicating a high degree of short range
order. 
We also estimated the `correlation length' $\xi$ 
associated with the 
domain structures, from a Lorenztian fit to the B-B$^{\prime}$ structure factor 
of the form
$S_{BB^{\prime}}({\bf q}) \sim
{\xi}^{-1}/((q_x - \pi)^2 + (q_y - \pi)^2 + \xi^{-2})$.
This yields 
 $\xi \sim 6.6,~5.9,~4.8,~3.6$.

Overall, the disorder in these systems 
should be characterised in terms of two variables: $S$, which is
a gross measure of order, and $p$ (or $\xi$) which quantify 
short range correlation. In general, physical properties would
depend on both of these and not simply $S$.

\subsection{Magnetic order with antisite disorder}

In the FM  case it is simple to see that the presence
of AF superexchange at the APB would tend to align spins in 
opposite directions across an antiphase boundary.
The system breaks up into up and down spin domains. 
Suppose the `up spin' domains
correspond to the correctly located sites and are the
`majority'.
The net magnetisation is 
proportional to the volume difference between the correctly
located and `mislocated' regions. If the degree of mislocation
is $x$, then the normalised magnetisation $M = (1-x) -x = 1-2x
=S$. In the magnetic structure factor $D({\bf q})$, 
the FM peak is at ${\bf Q}_F = \{0,0\}$. 
By definition $D({\bf Q}_F)  = M^2$, and 
from the domain argument
this is simply $S^2$.
This dependence is well established experimentally
\cite{theor-asd2}, and also observed by us \cite{vn-pm-fm}.

For the A type phase, studied on the same
antisite structures as the FM, the nature of local magnetic
order is more subtle. For the clean A-AF  
the order is at {\it two possible pairs}
of wavevectors,
${\bf Q}_{A1} = \{\pi/2, \pi/2\}$ and ${\bf Q}_{A2}
=\{3\pi/2, 3\pi/2\}$,
or ${\bf Q}_{A3} = \{\pi/2, 3\pi/2\}$ and 
${\bf Q}_{A4} = \{3\pi/2, \pi/2\}$. 
We have set the lattice spacing $a_0=1$ on the DP lattice.
The two sets arise due to the two possible
diagonals along which the FM stripes can order. Within each set
there are two ${\bf Q}$ values because half the sites in 
the DP lattice are non-magnetic, and the spin field has to
have nodes there. 
In the clean system either $\{ {\bf Q}_{A1},{\bf Q}_{A2} \}$ or
$\{ {\bf Q}_{A3},{\bf Q}_{A4} \}$ are picked. In a disordered
system all four can show up, as in the middle column
bottom row in Fig.3.
For the G type phase the order is at the single pair:
${\bf Q}_{G1} = \{\pi, 0\}$ and ${\bf Q}_{G2} = \{0, \pi\}$.

The middle and right columns in Fig.3 show A type and G type order,
respectively, for progressively increasing ASD. Fig.4 quantifies
the suppression of the ordering peak in $D({\bf q})$, after 
disorder averaging over copies with roughly fixed degree of
mislocation.

We will analyse the structure factor, $D({\bf q})$,
 in terms of the domain
pattern.
$D({\bf q})$ is related to the Fourier transform of
the spin configuration:
\begin{eqnarray}
D({\bf q}) & = &  {1 \over V^2} \vert {\vec f({\bf q})} \vert^2 \cr
&&\cr
{\vec f({\bf q})} & = & \sum_{\bf R_i} 
{\bf S}_i 
e^{i {\bf q}.{\bf R}_i} 
\nonumber
\end{eqnarray}
where $V$ is the total volume of the system.

For collinear order, where the spin  projection is only on the
$z$ axis, the spin vector can be replaced by $S_i$.
If we imagine the spin configuration to be broken up into
domains, indexed by a label $\alpha$, say, then, for
collinear phases, ${\vec f({\bf q})} = {\hat z} f({\bf q})$
and:
\begin{equation}
f({\bf q})  =  \sum_{\alpha} \sum_{{\bf R}_i^{\alpha}}
S_i e^{i {\bf q}.{\bf R}_i^{\alpha}}
= \sum_{\alpha} f_{\alpha}({\bf q})
\end{equation}
where the sum runs over the domains, and the ${\bf r}_i^{\alpha}$ are
coordinates within a domain. This shows that $f({\bf q})$ gets
additive contributions from various domains, with phase
factors that we will soon clarify. The formulation above holds
as long as (a)~there is no significant non collinearity and 
(b)~$\xi \gg 1$, {\it i.e.}, we can
ignore `interfacial' spins which may be hard to assign to any
particular domain.

\begin{figure}[t]
\centerline{
\includegraphics[width=4.4cm,height=5.0cm,angle=0]{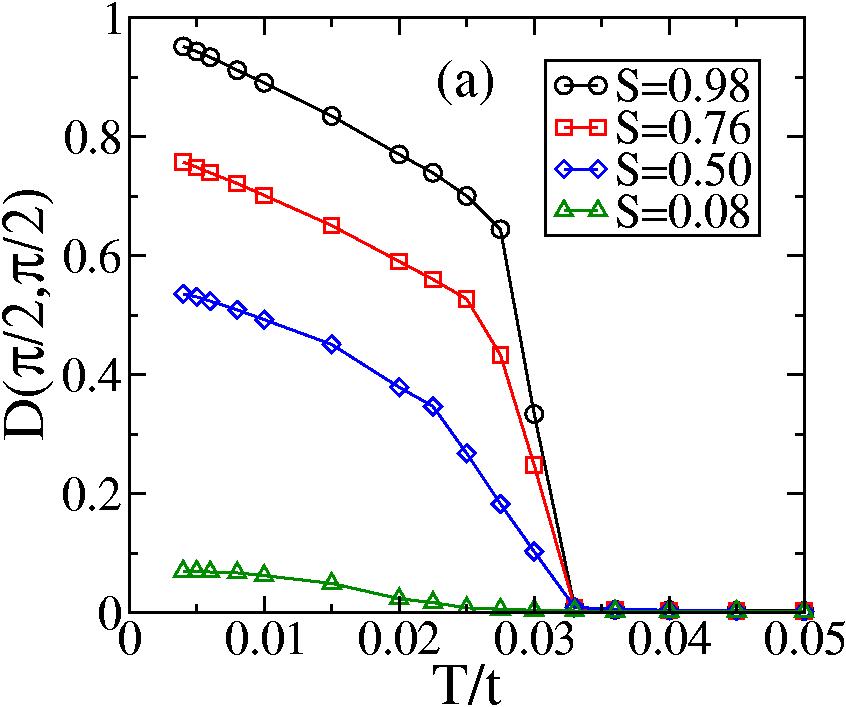}
\hspace{-.3cm}
\includegraphics[width=4.4cm,height=5.0cm,angle=0]{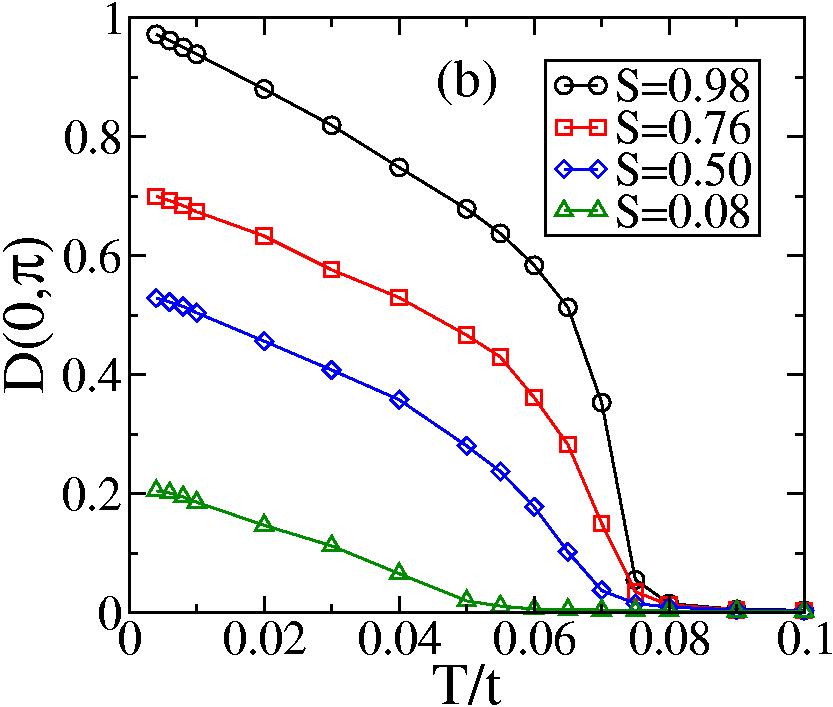}
}
\caption{Colour online: Magnetic order in the A type and G type
phases with ASD. The results are based on MC on $40 \times 40$
systems, and averaged typically over 10 configurations for each
value of $S$.}
\end{figure}

\subsubsection{A type order}

For weak to moderate  ASD we observe that 
the system prefers FM stripes along any single diagonal, albeit with
phase slippage between the stripes to accommodate the effect
of $J_{AF}$. This is true for $S=0.98,~0.76$ and $0.50$, 
the top three
rows in the middle column in Fig.3. Domains which are
translated  with respect to the reference domain have
relative displacement 
$\delta {\bf r}_{\alpha} = {\hat x} a_0$ or ${\hat y} a_0$.
The order {\it within} all domains is similar. So,
the contribution of each domain 
at the ordering wavevector, ${\bf Q}$,
 will be proportional
to the domain volume, and involve a phase factor:
$$
f_{\alpha}({\bf Q}) = V_{\alpha} e^{i {\bf Q}. {\delta} {\bf r}_{\alpha}}
f_0({\bf Q})
$$
where $f_0({\bf Q})$ is the normalised reflection in the perfectly 
ordered system. 
For the ${\bf Q} = {\bf Q}_{A1} = 
\{\pi/2, \pi/2\}$ peak, the phase factor is $e^{i \pi/2}$ 
irrespective of whether the domain 
is $x$ displaced or $y$ displaced.
So, all the `mislocated' sites, grouped into domains, contribute
$V_{mis} e^{i \pi/2} f_0({\bf Q})$, where $V_{mis}$ is the total
volume of mislocated regions.

Adding the contribution from the majority domains, which are
undisplaced, we obtain:
$$
f({\bf Q}_{A1}) = ((V - V_{mis})  + V_{mis} e^{i \pi/2} )f_0({\bf Q}_{A1})
$$
Remembering that $V_{mis}/V = x$, the volume normalised
structure factor peak is
$$
D({\bf Q}_{A1}) = \vert (1-x) +  e^{i \pi/2}x \vert^2 = 
{1 \over 2} (1 + S^2)
$$
This is roughly consistent with the $S$ dependence of
the $T \rightarrow 0$ structure factor in Fig.4.(a). It is 
{\it distinctly slower} than the suppression of order
in the ferromagnet, where $D({\bf Q}_F) \sim S^2$.

At larger ASD however, the system has short stripes 
oriented along {\it both} 
diagonals, see bottom row
middle panel in Fig.3. These domains have magnetic peaks at 
${\bf Q}_{A3},{\bf Q}_{A4}$ and not at
${\bf Q}_{A1},{\bf Q}_{A2}$. Even assuming that the
`majority' domains all contribute $(1 -x) f_0({\bf Q}_{A1})$,
we notice that the mislocated regions require classification
into two groups: 
those contributing  at ${\bf Q}_{A1},{\bf Q}_{A2}$ with 
volume fraction $y$, say,  
and those at ${\bf Q}_{A3},{\bf Q}_{A4}$ with volume fraction
$x-y$.  
In that case the peak at ${\bf Q}_{A1}$ would be
$$
D({\bf Q}_{A1}) = \vert (1-x) +  e^{i \pi/2}y \vert^2 
$$
The $(x-y)$ fraction makes no contribution to the peak, and that
weight is `lost'. Notice that $ x \ge y \ge 0$, and it is not
possible to write $D({\bf Q}_{A1})$ purely in terms of $S$.
We could write the expressions for the structure factor 
at the other
three ${\bf Q}$ as well, and they will all depend on both
$x$ and $y$. This is a general feature of magnetic states 
where the
order can locally pick out different orientations.

We can make some headway in the strong disorder limit,
$x=1/2$, $S=0$, by assuming that there are four kinds of 
domains, with roughly equal area. There would be two 
families of $\{ {\bf Q}_{A1},{\bf Q}_{A2} \}$ domains, each with
$1/4$ the system volume, and a relative phase shift  $\pi/2$.
Similarly there would be two families of $\{ {\bf Q}_{A3},{\bf Q}_{A4} \}$
domains, each with volume $1/4$, and relative phase shift  $\pi/2$.
In this case the
${\bf Q}_{A1}$ peak, for example, in D({\bf Q}) will be:
$$
D({\bf Q}_{A1}) = \vert 1/4  +  e^{i \pi/2}/4  \vert^2 =1/8
$$
This is not very far from $D({\bf Q}_{A1}) \sim 0.1$ that we
obtain from the Monte Carlo. Other peaks, at 
${\bf Q}_{A2},{\bf Q}_{A3}$, {\it etc},
would have similar magnitude.

\subsubsection{G type order}

G type order occurs at the combination $\{ {\bf Q}_{G1},{\bf Q}_{G2}\}:$
$\{ \{0, \pi\},~\{\pi, 0\} \}$.
As before the relative displacement of the domains can be only
${\hat x} a_0$ or ${\hat y} a_0$. Suppose we are computing
the structure factor at ${\bf Q}_{G1}$ then all domains will
contribute, but with following phase factors: zero if the domain
is not mislocated ($\delta {\bf r} =0$), zero again if
the domain is ${\hat x}$ displaced, and $e^{i\pi} =-1$ if the domain
is ${\hat y}$ displaced. 

\begin{figure}[b]
\centerline{
\includegraphics[width=4.4cm,height=5.0cm,angle=0]{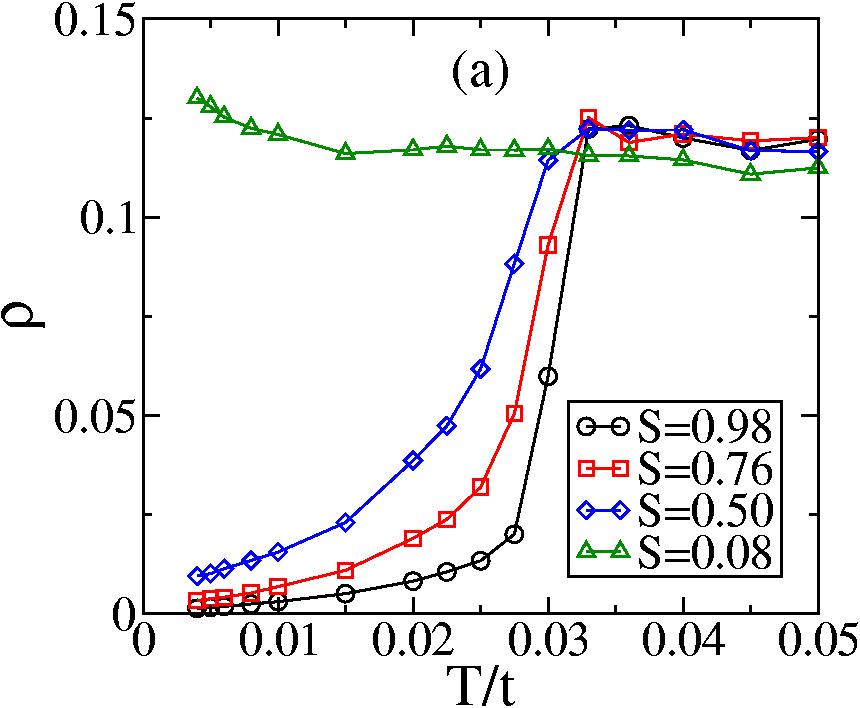}
\hspace{-.1cm}
\includegraphics[width=4.4cm,height=5.0cm,angle=0]{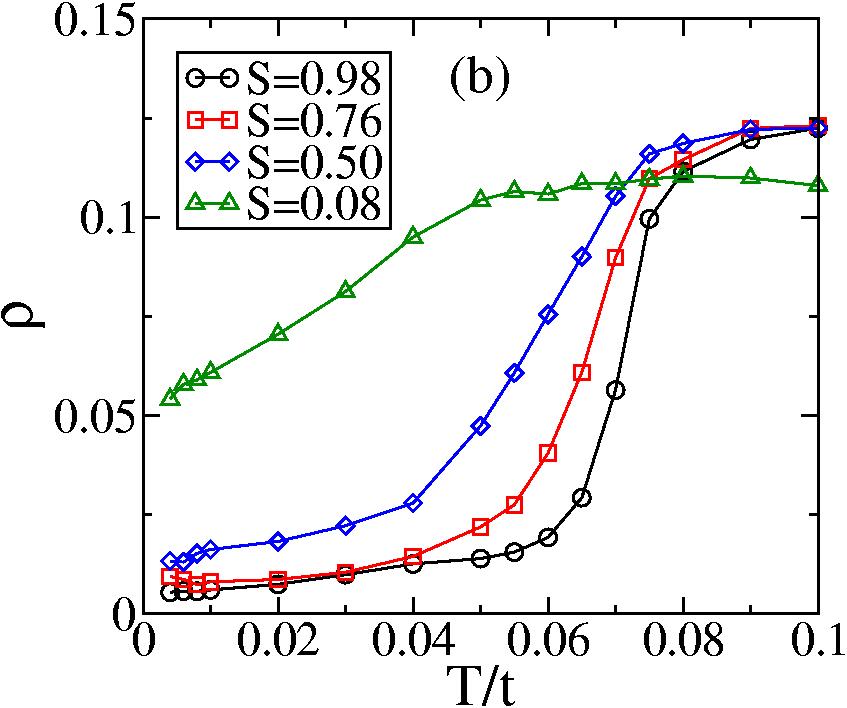}
}
\caption{Colour online: Resistivity in the A type and G type phases with ASD.
The results for each $S$ are averaged over 10 realisations of disorder.}
\end{figure}
\begin{figure}[t]
\centerline{
\includegraphics[width=4.8cm,height=4.7cm,angle=0]{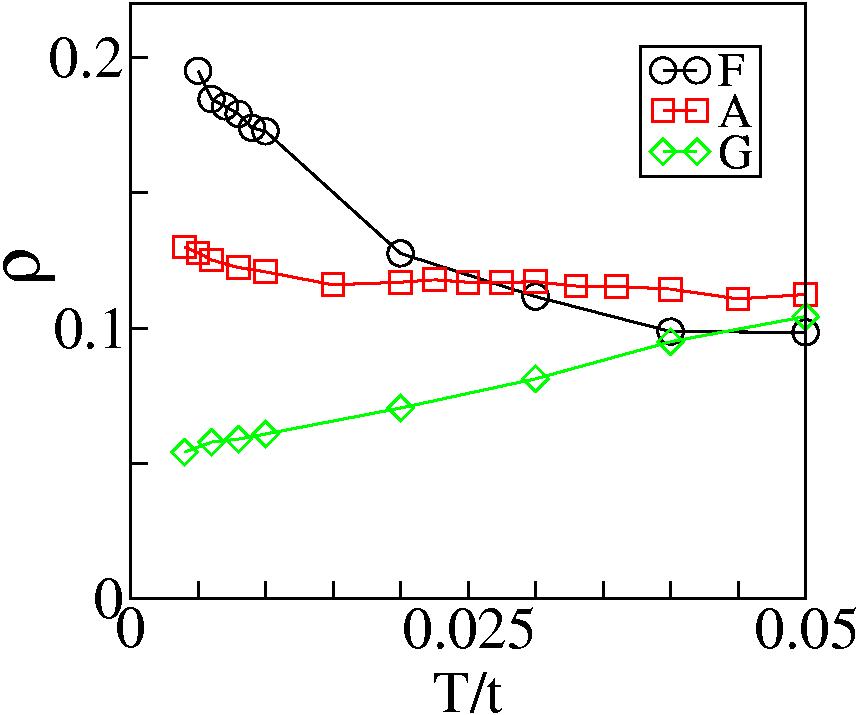}
}
\caption{Colour online: Comparison of the resistivity in the FM, A and G type
correlated phases in samples with the lowest degree of order $S=0.08$.}
\end{figure}

In {\it two domain} systems, 
as in the second row in Fig.3, 
the mislocated domain is either  $x$ displaced or
$y$ displaced. For copies with $x$ displacement the 
contribution 
at ${\bf Q}_{G1}$ will be $\vert (1-x) + x\vert^2 =1$, while
for $y$ displacement it will be $\vert (1-x) + e^{i\pi} x\vert^2 
=( 1 - 2x)^2 = S^2$.  Averaging the {\it structure factor}
 over copies would lead to 
$ D({\bf Q}_{G1}) = (1/2) (1 + S^2) $.
This is roughly what we observe in our Fig.4.(b) at $T=0$. 

In large systems, where there will be many domains, we can
assume that half the mislocated domains are $x$ displaced
and half $y$ displaced. In that case the structure factor
would be 
$$
D({\bf Q}_{G1}) = \vert (1-x) + x/2 + e^{i\pi} x/2\vert^2
$$
Using $1 -2x = S$ this leads to
$$D({\bf Q}_{G1})
= (1/4)(1 + S)^2
$$
For $S \rightarrow 0$ this gives $0.25$, not far from $\sim 0.20$
that we obtain from our configurations. 

This reveals that for both A and G type order 
even when {\it half the sites are mislocated}, {\it i.e}, one has
maximal antisite disorder, there is a surviving peak in the
structure factor. All these of course assumed that the structural
pattern had a high degree of spatial correlation so that
one can meaningfully talk of domains. We should have $ 1 - p \ll 1$,
or structural correlation length $\xi \gg a_0$. If the
structures were fragmented to a random alloy then the results
above would not hold. We have checked this explicitly.

\subsection{Transport with antisite disorder}

All the three phases, FM, A type, and G type, in the 2D double
perovskite model are metallic in the clean limit.
The electronic states are extended, and there is 
a finite density of states at the  Fermi level.  In the absence
of antisite disorder the resistivity, $\rho(T)$, in
all three have similar temperature dependence. The resistivity
increases rapidly as $T$ increases towards $T_c$, Fig.5 and our
earlier work \cite{vn-pm-fm} on the FM, 
and `saturates' at high temperature.
A Fisher-Langer type \cite{fisher-langer}
 phenomenology can qualitatively describe 
the transport.

Weak disorder leads to an increase in the residual 
resistivity of the A and G type phases, see Fig.5,
  as observed earlier for the FM. The sharp resistive
transition observed in the clean limit $(S=0.98)$ is
also gradually broadened with increasing disorder. 
There is however a key 
difference with respect to the FM phase when we move to
strong disorder.

In the 2D case our results \cite{vn-pm-fm} on the FM
suggest 
an insulating $T=0$ state beyond a 
critical disorder, with $d\rho/dT <0$. 
At the highest disorder, $S=0.08$, 
$\rho(T)$ in the A type AF remains essentially flat
down to $T=0$, while in the G type 
phase 
there is still a low temperature {\it downturn}. 
While these are finite size results,
we argue below why there is an intrinsic reason for transport in the AF phases
to be less sensitive to antisite disorder and domain formation.
This is related to the nature of electronic wavefunctions in 
these phases. 
We consider the two phases in succession below.

\begin{figure*}
\centerline{
\includegraphics[width=5.6cm,height=2.8cm,angle=0]{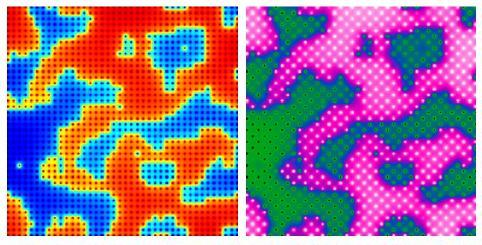}
\hspace{-.3cm}
\includegraphics[width=5.6cm,height=2.8cm,angle=0]{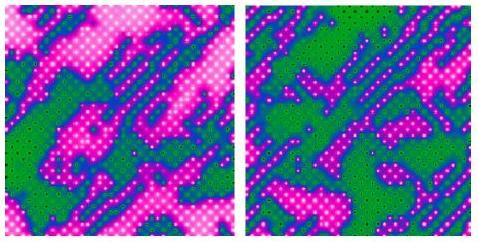}
\hspace{-.3cm}
\includegraphics[width=5.6cm,height=2.8cm,angle=0]{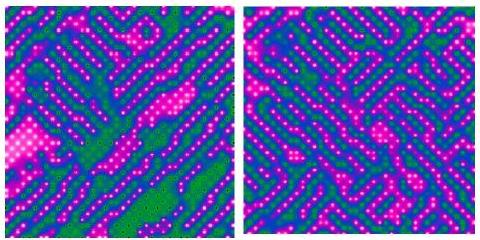}
}
\caption{Colour online: Magnetic correlations with increasing electron
density as one traverses the coexistence regime in an antisite
disordered background.  The first panel shows
the structural pattern arising from the ASD. Panel 2 shows
the spin correlations at $\mu = \mu_{FM} = -1.8$, where the ground state is
a (domain) ferromagnet. The extreme right is for $\mu = \mu_{AF}
= -1.4$, where the system has only A type AF correlations. As
$\mu $ increases from $\mu_{FM}$ to $\mu_{AF}$ the pattern
exhibits coexistence of short range FM and AF correlations.}
\end{figure*}

In the A type phase the core spin order involves diagonal stripes.
`Up-spin' electrons delocalise on down-spin stripes which involve 
one B diagonal and the two adjacent B$^{\prime}$ lines. `Down-spin' electrons 
delocalise on up-spin stripes. The down and up stripes 
share a common B$^{\prime}$ line. The
essential feature is that the electronic wavefunctions are
quasi-one dimensional. The introduction of antisite
disorder  leads to two effects: (i)~it hinders  propagation along
the stripe, and (ii)~allows scattering between the stripes
leading to an `expansion' of the wavefunction 
in the  transverse direction.
On its own, the first effect would have suppressed conduction,
but the new matrix element between stripes allows a transverse
pathway for delocalisation.
In contrast to the FM case there is no `confinement' of the
wavefunctions to specific domains, and the two competing effects
above lead to a finite resistivity (at least a much weaker
upturn)  even at strong disorder. 

\begin{figure}[b]
\centerline{
\includegraphics[width=5.0cm,height=5.0cm,angle=0]{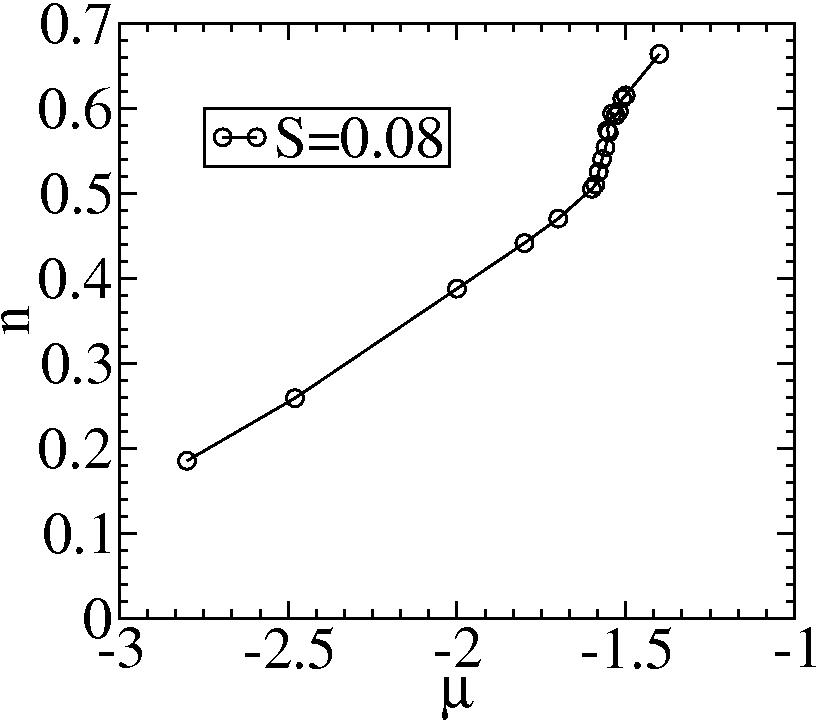}
}
\caption{Colour online: The variation of electron demsity with
chemical potential as the FM to A type AF crossover is traversed
in a high ASD sample ($S=0.08$).
}
\end{figure}

In the clean G type phase the system can be viewed as
two interpenetrating
square lattices, one with up spin B sites, the other with down
spin B sites. The electrons delocalise via the B$^{\prime}$ sites, and 
each B$^{\prime}$ site hosts both up and down electron states.  Electrons 
in both spin channels are delocalised over the whole system
in the absence of ASD. The presence of antisite domains 
leads to scattering but no confinement of electrons to the
domains. Due to the inherently 2D character of the G type
electronic states, in contrast to quasi-1D for A type, the
system has a lower resistivity.

We contrast the resistivity of the three phases in Fig.6 for
the case of maximum disorder that we have studied. These are
configurations with $S \sim 0.08$, but, as we have noted, with
a fairly high degree of local correlation. The FM has a clear
low $T$ upturn due to the confinement of electrons into
domains or pathways created by the ASD. The A and G type 
electronic states, in either spin channel, are {\it not confined} 
to the magnetic/structural domains, and the resistivity
remains comparatively lower.
The high temperature resistivity is 
determined by spin disorder scattering, depends  weakly on
carrier density (the FM, A and G phases have different $n$), and
is almost temperature independent.

\subsection{Phase coexistence}

In the clean limit, the increase of electron density by doping the
FM would encounter a window of phase separation. A homogeneous
state is not allowed for $ 0.45 \le n \le 0.58$ 
and this (idealised)
system would break up into macroscopic regions having densities
$n \sim 0.45$  and $n \sim 0.58$. 
This pathology is avoided by long range Coulomb
interactions or quenched disorder. For the antisite disordered
configurations that we are considering the ASD itself 
controls the pattern of spatial coexistence. 

Fig.7 shows the magnetic correlations in a fixed antisite background
for changing electron density. The leftmost panel is the structural
pattern, showing the antisite domains. The next five snapshots
correspond to increasing chemical potential, $\mu$, and consequently
the electron density. The first panel in this set is a 
(domain) ferromagnet at $n < 0.45$, the second shows emergence of
stripes along with the FM regions. The FM regions shrink and
the linelike patterns become more prominent in the third panel. 
The fourth and fifth panel complete the evolution, with FM
correlations completely replaced by stripes (of both orientation)
as in the bottom of middle column in Fig.3. 
The evolution of the particle density with $\mu$, and the 
rapid change near the phase separation window, are shown in
Fig.8. 


\section{Conclusions}
We have studied the survival of the antiferromagnetic 
double perovskite phases in the presence of spatially correlated
antisite disorder. 
We observe that antisite disorder affects the antiferromagnetic 
order  much less strongly than it affects ferromagnetism.
For a given structural order parameter $S$, the 
A type AF structure 
factor follows $D_A \sim (1 + S^2)/2$, in contrast
to $D_F \sim S^2$ in the ferromagnet, while the
G type phase follows $D_G \sim (1 + S)^2/4$.
So, despite the possibility
of large antisite disorder at the high electron doping
needed to observe the AF phases, there is certainly hope
of observing these magnetic structures.
The AF states are metallic, and the electronic wavefunctions in
these phases continue to be spatially extended even at large
disorder. Antisite disorder increases the residual resistivity,
but, unlike the ferromagnet, we 
did not observe any insulating regime.
The field response of these AF metals is also fascinating,
and will be separately discussed.

{\it Acknowledgements:}
We thank T. Saha Dasgupta, P. Sanyal and S. Ray for early discussions
and Rajarshi Tiwari for collaboration on related issues.
We acknowledge use of the Beowulf Cluster at HRI.
PM acknowledges support from a DAE-SRC Outstanding 
Research Investigator
Award, and the DST India through the Indo-EU ATHENA project.


\begin{thebibliography}{99}
\bibitem{footnote-b-bprime} 
This refers to a situation where the effective
level difference between B and B$^{\prime}$ is zero. 
If this difference
were comparable to $t$ one would begin to see additional
`potential scattering' effects arising from the antiphase boundaries,
enhancing the overall resistivity.
The present study focuses purely on how magnetic order affects transport.
\end{thebibliography}
\end{document}